\documentclass[12pt]{iopart}

\usepackage{iopams}
\usepackage{setspace}

\usepackage{amsfonts}   
\usepackage{amssymb}
\usepackage{graphicx,epstopdf}   

\begin{document}

\title{On an exactly solvable toy model and its dynamics}

\author{Yang K L$^{1, 2}$ and Zhang J M$^{1,2}$}
\address{$^1$ Fujian Provincial Key Laboratory of Quantum Manipulation and New Energy Materials,
College of Physics and Energy, Fujian Normal University, Fuzhou 350007, China}
\address{$^2$ Fujian Provincial Collaborative Innovation Center for Optoelectronic Semiconductors and Efficient Devices, Xiamen 361005, China}

%
%
%

\begin{abstract}
The eigenstates and eigenenergies of a toy model, which arose in idealizing a local quenched tight-binding model in a previous publication [Zhang and Yang, EPL \textbf{114}, 60001 (2016)], are solved analytically. This enables us to study its dynamics in a different way. This model can serve as a good exercise in quantum mechanics at the undergraduate level.
\end{abstract}

\maketitle

\section{Introduction}

Exactly solvable models are rare, and it is especially the case when accessibility to undergraduate students is required. In a typical undergraduate quantum mechanics course, the few exactly solvable models mentioned are often limited to the infinite square well potential, the hydrogen atom, the harmonic oscillator, and a charged particle in a uniform magnetic field. But it is definitely a good idea to have more such models in store and use them as exercises in teaching. From this point of view, it is a shame that the pedagogical value of some simple models in research articles \cite{inverse, weber, zhangprl,longhi} has not yet been fully appreciated.

In this paper, we would like to promote the awareness of a recently devised toy model, which is exactly solvable and can make a good exercise for students. The model was motivated by the study of the quench dynamics of a Bloch state on a tight-binding model \cite{epl1,epl2,prb} and accounts for the peculiarities of the dynamics very well. Here we detach it from the original context and formulate it in general terms. It consists of infinitely many levels $\{|n \rangle , n\in \mathbb{Z} \}$, and its Hamiltonian is
$
  H  = H_0 + H_1  $,
where the diagonal part $ H_0$ is
\begin{eqnarray}\label{H0}
  H_0 &=& \sum_{n = -\infty}^\infty n \Delta |n \rangle \langle n |,
\end{eqnarray}
which means that the levels are equally spaced with the gap between two adjacent levels being $\Delta$;
and the off-diagonal part $H_1$ is
\begin{eqnarray}\label{H1}
  H_1 &=& \sum_{n_1,n_2 = -\infty}^\infty  g |n_1 \rangle \langle n_2 | ,
\end{eqnarray}
which means that two arbitrary levels, regardless of their energy difference, are coupled to each other with constant strength $g$. For the Hamiltonian to be hermitian, the two parameters $\Delta $ and $g$ of course should be real.

For those who know the Stey-Gibberd model \cite{stey,ejp}, which is important for modeling quantum decay and quantum resonance \cite{glushkov1, moiseyev, glushkov2, glushkov3, glushkov4}, the present model might look very similar. However, the difference is substantial. The Stey-Gibberd model is of the Hamiltonian
\begin{eqnarray}
  H_{SG} &=&  \sum_{n = -\infty}^\infty n \Delta |n \rangle \langle n | + E_d |d \rangle \langle d | + g \sum_{n = -\infty}^\infty(|d \rangle \langle n | + |n \rangle \langle d | ) .
\end{eqnarray}
It consists of an equidistant quasi-continuum $\{|n \rangle , n\in \mathbb{Z} \}$  plus a defect mode $|d \rangle $, while here we have only the quasi-continuum. Moreover, in the Stey-Gibberd model, there is no coupling between the levels belonging to the quasi-continuum band, while here any two of them are coupled.

\section{Stationary properties}
While the authors of Refs.~\cite{epl1,epl2,prb} studied the dynamics of the model, they did that by solving the time-dependent Schr\"odinger equation directly without solving the eigenstates and eigenenergies of the model. It is our purpose to fill this gap below.

Even without calculation, we can see that if $
  |\psi\rangle  = \sum_{n=-\infty }^\infty a_n |n\rangle
$ is an eigenstate of $H$ with eigenenergy $E$, the translated state
$  | \tilde{\psi}\rangle  = \sum_{n=-\infty }^\infty a_n |n+1\rangle $
is an eigenstate with energy $E + \Delta$. Formally, we can introduce the translation (raising) operator
$   T =   \sum_{n = -\infty}^\infty  |n+1 \rangle \langle n | $.
It is then readily checked that $  [H_0, T] = \Delta  T $ and $[H_1, T] = 0 $.
Hence
\begin{eqnarray}
  H T |\psi\rangle  &=& (T H_0 + \Delta T + T H_1) |\psi \rangle = ( \Delta T + T H) |\psi \rangle = (E + \Delta )T |\psi \rangle ,
\end{eqnarray}
which means the state $|\tilde{\psi} \rangle = T | \psi \rangle $ is an eigenstate of $H $ with eigenenergy $ E + \Delta $.
Therefore, the eigenenergies are equally spaced too.

To determine their precise values, we project the two sides of the equation $H|\psi \rangle = E | \psi \rangle $ onto the basis state $| n \rangle $.
We get $
  E a_n  = n \Delta a_n + g S   $,
with
\begin{eqnarray}\label{defS}
  S &=& \sum_{n=-\infty}^\infty a_n.
\end{eqnarray}
The observation is that $S$ is independent of $n$. We solve formally
\begin{eqnarray}\label{an}
  a_n &=& \frac{g S}{E - n \Delta } .
\end{eqnarray}
When this expression is substituted back into (\ref{defS}), the factor $S$ drops out, and we get an equation for $E$,
\begin{eqnarray}\label{unity}
  1 &=& \sum_{n=-\infty}^\infty \frac{g }{E - n \Delta } .
\end{eqnarray}
By the famous formula \cite{aigner}
\begin{eqnarray}\label{formula1}
 \pi \cot \pi x  &=& \sum_{n=-\infty}^\infty \frac{1}{x- n },\quad x \notin \mathbb{Z} ,
\end{eqnarray}
we then reduces (\ref{unity}) to $1 = (\pi g/\Delta) \cot (\pi E /\Delta )$, from which we solve finally the  eigenenergies as ($-\infty < m < \infty $)
\begin{eqnarray}
  E_m &=& \Delta \left(m + \frac{1}{\pi } \arctan \frac{\pi g}{\Delta} \right) .
\end{eqnarray}
Indexed by the integer $m$, they are indeed equally spaced by $\Delta $. It is convenient to introduce the variable
\begin{eqnarray}
  \alpha  &=&  \frac{1}{\pi } \arctan \frac{\pi g}{\Delta},
\end{eqnarray}
which takes value in the interval $(-\frac{1}{2}, \frac{1}{2})$, so that we can write $E_m = \Delta (m + \alpha )$.

We have yet to determine the corresponding eigenstate $|\psi_m \rangle $, i.e., the coefficients $a_n$. By (\ref{an}), we know $a_n = C /(E_m /\Delta - n ) =C /(m-n + \alpha  )$. The normalization condition requires
\begin{eqnarray}
  1 &=& \sum_{n=-\infty}^\infty |a_n|^2 = |C |^2 \sum_{n=-\infty}^\infty \frac{1}{(m - n +\alpha )^2 } .
\end{eqnarray}
By the formula
\begin{eqnarray}
  \left( \frac{\pi}{\sin \pi x }\right)^2 &=& \sum_{n=-\infty}^\infty \frac{1}{(x- n )^2},
\end{eqnarray}
which can be obtained from (\ref{formula1}) by differentiation with respect to $x $, we solve
\begin{eqnarray}
  |C |^2 &=& \frac{(\pi g/\Delta )^2}{\pi^2 [1 + (\pi g/\Delta)^2] } =\frac{\sin^2 \pi \alpha }{\pi^2}   ,
\end{eqnarray}
which is independent of the state index $m$. This finishes the solution of the eigenvalues and eigenstates.

\section{A brief digression}
One might wonder what characteristic of the model is responsible for its exact solvability. We would say that it is the rank-one property of the perturbation $H_1 $. Introducing the state $|\phi \rangle = \sum_{n\in \mathbb{Z}} |n \rangle $, we can write $H_1$ succinctly as $H_1 = g |\phi \rangle \langle \phi | $. It is then clear that the matrix corresponding to $H_1$ is of rank one, the lowest rank possible for a nonzero matrix. Noting that in problems with matrices and integral equations, rank-one perturbations \cite{ding} and rank-one kernels \cite{tricomi}, respectively, often allow simple analysis, here one can also anticipate some simplicity.

Let us consider a more general model. Let $J $ be an arbitrary subset of $\mathbb{Z}$, and let the Hamiltonians be $ {H}_0 = \sum_{n\in J } \omega_n |n \rangle \langle n | $ and $ {H}_1 = g |\varphi \rangle \langle \varphi | $. Here the energies $\omega_n $ of the levels $|n \rangle $ are arbitrary (but real of course) and the state $|\varphi \rangle = \sum_{n\in J } c_n |n \rangle $ is also arbitrary. In comparison with (\ref{H0}) and (\ref{H1}), here we have dropped some specialities of $H_{0,1} $ but have retained the rank-one feature of $H_1$.

Consider the eigenvalue equation $( {H}_0 + {H}_1)|\psi \rangle = E | \psi \rangle $. Projecting the two sides onto $|n \rangle$, we get
$
  E \langle n | \psi \rangle  = \omega_n \langle n | \psi \rangle + g \langle n | \varphi \rangle \langle \varphi | \psi \rangle
$, from which we solve
\begin{eqnarray}
  \langle n | \psi \rangle  &=& \frac{g c_n }{E - \omega_n } \langle \varphi |\psi \rangle .
\end{eqnarray}
But $\langle \varphi |\psi \rangle = \sum_{n\in J } \langle \varphi | n \rangle \langle n | \psi \rangle = \sum_{n\in J } c_n^* \langle n | \psi \rangle $, hence
\begin{eqnarray}
  \langle \varphi |\psi \rangle &=&  \sum_{n\in J } \frac{g |c_n|^2 }{E - \omega_n } \langle \varphi |\psi \rangle .
\end{eqnarray}
Factoring out the unknown $ \langle \varphi |\psi \rangle $, we get the equation for the eigenvalue $E$,
\begin{eqnarray}
  1 &=&\sum_{n\in J } \frac{g |c_n|^2 }{E - \omega_n } .
\end{eqnarray}
This equation is apparently simpler, both analytically and numerically, than what one would obtain from the condition $\det (E - {H}_0 - {H}_1) = 0 $ in case $H_1$ lacks the rank-one property. Further simplification occurs as in (\ref{unity}) when $J = \mathbb{Z}$, $\omega_n = n \Delta$, and $c_n \equiv 1 $.

Finally, we would like to point out that at least in one dimension, a rank-one perturbation can be readily realized with a delta potential. Actually, this is how it is realized in the problem in Refs.~\cite{epl1,epl2,prb}. Suppose $H_0= -\frac{\partial^2 }{\partial x^2 } +V(x)$ is the unperturbed Hamiltonian of a particle in a one-dimensional potential $V(x)$, whose eigenstates $\{  \chi_{n\geq 1 } (x) \}$ are all normalizable. Now turn on a delta potential at $x_0$, i.e., add a potential $H_1 = g \delta(x- x_0 )$ to $V(x)$. In the basis of $\{  \chi_{n\geq 1 } (x) \}$, $H_0$ is diagonal, while the matrix elements of $H_1$ are
\begin{eqnarray}
  \langle \chi_m| H_1 | \chi_n \rangle &=& g\int_{-\infty}^{+\infty} dx \chi_m^*(x) \delta(x-x_0) \chi_n(x) = g \chi_m^*(x_0)\chi_n(x_0) ,
\end{eqnarray}
which apparently correspond to a rank-one matrix.

\section{Dynamics}

Having obtained the eigenstates and eigenvalues explicitly, we can then use them to solve dynamical problems. For example, let us revisit the quench problem in Refs.~\cite{epl1,epl2,prb}, which was solved in a different approach. Suppose initially the system is in the level  $ |0 \rangle $. As time evolves, due to the coupling $H_1$, all other levels will be populated too. A quantity of primary concern is the probability of transition into an arbitrary level $|n\rangle $. That is, we are interested in the matrix element of the time evolution operator ($\hbar =1 $)
\begin{eqnarray}
  A_n(t) &=& \langle n | e^{-i H t }| 0 \rangle ,
\end{eqnarray}
whose magnitude squared is the transition probability from $|0\rangle $ to $|n \rangle $ (or survival probability when $n=0 $).

Inserting the identity operator $I = \sum_{m\in \mathbb{Z} } |\psi_m \rangle \langle \psi_m | $, we get
\begin{eqnarray}\label{An}
  A_n(t) &=&  \sum_{m\in \mathbb{Z} } \langle n | \psi_m \rangle \langle \psi_m |0 \rangle  e^{-i E_m t }  \nonumber \\
  &=& \sum_{m\in \mathbb{Z} } \frac{|C|^2}{(m - n + \alpha  )(m + \alpha ) } e^{-i  (m + \alpha ) \Delta t}.
\end{eqnarray}
It is easily seen that
\begin{eqnarray}\label{period}
  A_n(t+ T ) = A_n(t) e^{-i 2 \pi \alpha  } ,
\end{eqnarray}
where $T= 2\pi/\Delta  $.  The problem then reduces to determining $A_n (t)$ for $0\leq t \leq T $. For this, we introduce the function
\begin{eqnarray}
  B(x;\alpha) &=& \sum_{m\in \mathbb{Z}} \frac{e^{-im x }}{m+\alpha } ,
\end{eqnarray}
for $\alpha \notin \mathbb{Z}$ and $0< x < 2 \pi $. In terms of $B$, we have
\begin{eqnarray}
  A_{n\neq 0 }(t) &=& |C|^2  e^{-i \alpha \Delta t }  \frac{e^{-in \Delta t} -1}{n} B(\Delta t; \alpha ) , \label{AinB1}  \\
  A_0(t) &=&   -|C|^2  e^{-i \alpha \Delta t }  \frac{\partial}{\partial \alpha } B(\Delta t; \alpha ) . \label{AinB2}
\end{eqnarray}
Now by considering the Fourier series of $e^{i \alpha x }$, $0\leq x \leq 2 \pi $, we have
\begin{eqnarray}
  B(x;\alpha ) &=& \frac{2 \pi i  }{e^{i 2\pi \alpha } -1}e^{i \alpha x } , \quad 0 < x < 2 \pi .
\end{eqnarray}
Substituting this explicit expression into (\ref{AinB1}) and (\ref{AinB2}), after some straightforward calculation, we get
\begin{eqnarray}
  A_{n\neq 0 }(t) &=& \frac{g}{n ( \Delta + i \pi g )}(e^{- i n \Delta t }-1), \label{Annot0} \\
  A_{ 0 }(t) &=& 1 - \frac{i g \Delta t }{\Delta + i \pi g  } ,\label{A0}
\end{eqnarray}
for $0\leq t \leq T$. Their values at an arbitrary time can then be determined by using (\ref{period}).

\begin{figure*}[tb]
\includegraphics[width= 0.351\textwidth ]{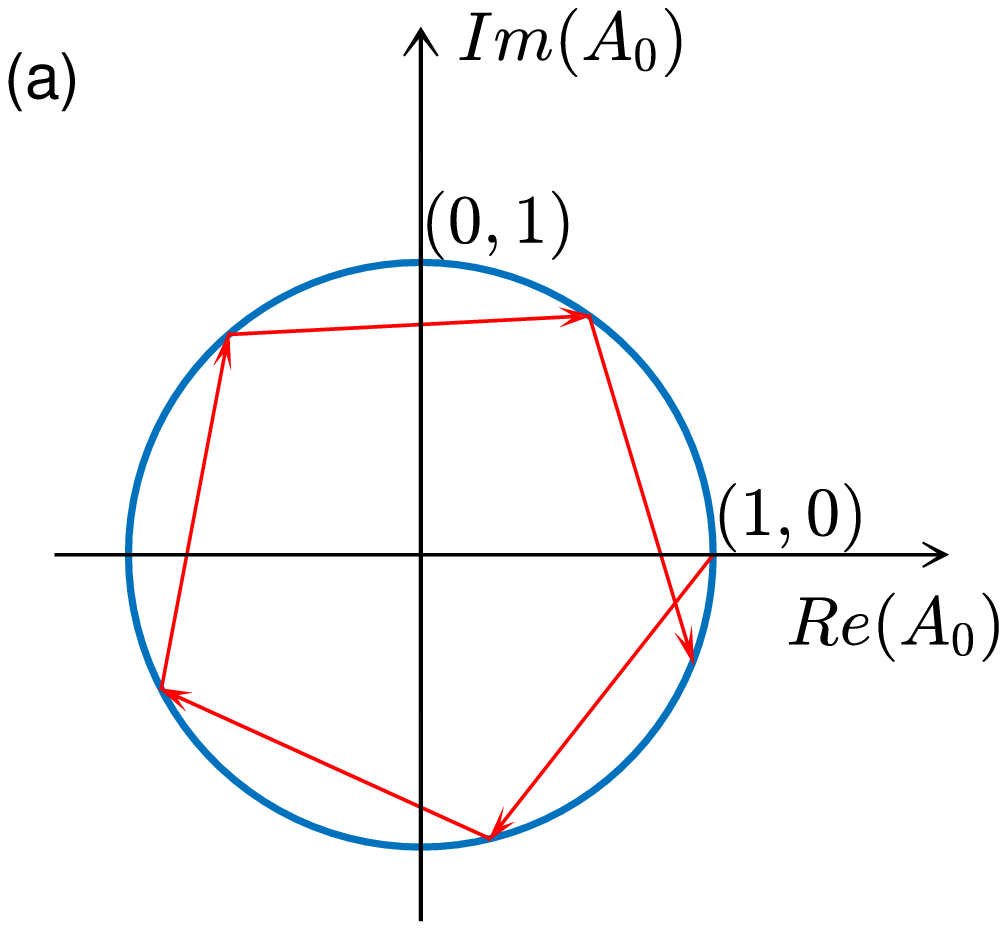}
\hspace{11mm}
\includegraphics[width= 0.41\textwidth ]{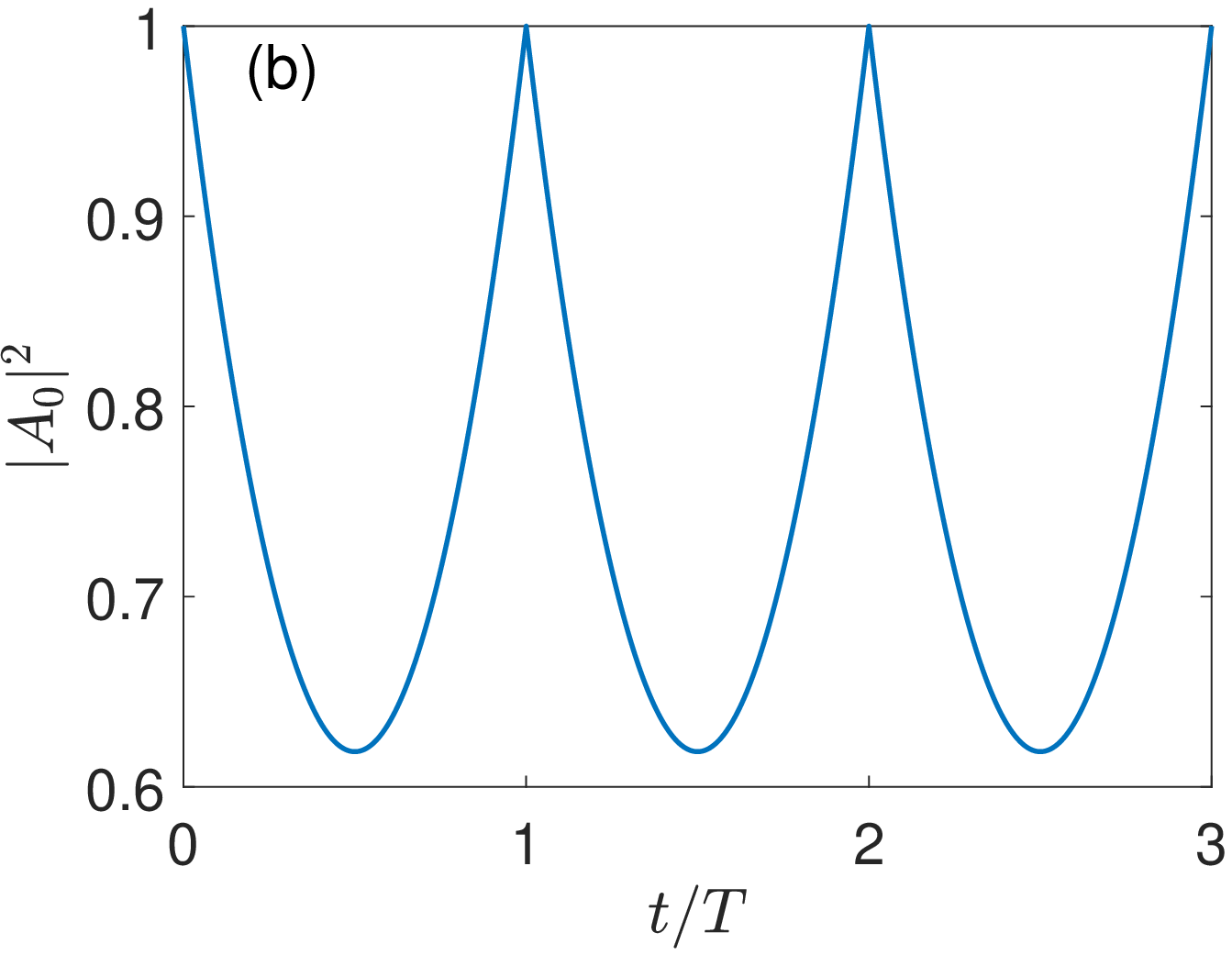}
\caption{(Color online) (a) Trajectory of the survival amplitude $A_0$ in the complex plane; (b) Time evolution of the survival probability $|A_0|^2$. The parameters are $\Delta = 1 $, $g= 0.25$. }
\label{fig1}
\end{figure*}

We note that $A_0(t)$ is a linear function of time $t$ when $0\leq t \leq T$. In this interval, it interpolates linearly between the end values $A_0(t=0) =1$ and $A_0(t=T ) = (\Delta - i \pi g )/(\Delta + i \pi g ) =e^{-i 2\pi \alpha }$, whose absolute values are both unity. This means that after time $T$, the system revives completely---The probability initially flew to other levels has all flown back. Indeed, by (\ref{Annot0}), $A_{n\neq 0 }(t = T ) =0 $. Apparently, like that in a harmonic oscillator, the perfect revival here is a consequence of the equidistant spectrum of the model.

Together with (\ref{period}), the fact that $A_0$ interpolates linearly between $1$ and $e^{-i2\pi \alpha }$ in the first period of $[0,T]$ means that it will do so between $e^{-i2\pi \alpha }$ and $e^{-i4\pi \alpha }$ in the second period of $[T, 2T]$, and so on. Overall, the picture is that $A_0$ follows a trajectory in the complex plane like a ball does in a circular billiard. See figure~1(a) for illustration.

The survival probability $|A_0|^2$  is then a periodic function of time which returns to unity when $t$ is an integral multiplier of $T$. In the first period, it has the explicit expression of
\begin{eqnarray}\label{A0squared}
  |A_0(t)|^2 &=& 1 - \frac{g^2 \Delta^2}{\Delta^2 + \pi^2 g^2 } t(T-t) ,
\end{eqnarray}
which is a polynomial of order two. The curve of $|A_0|^2$ is illustrated in figure~1(b). Note that whenever it returns to unity, it shows a cusp.

With (\ref{Annot0}), (\ref{A0}),  and (\ref{A0squared}), we have reproduced the main results of \cite{epl1}. Technically, the derivation here is not necessarily simpler; but conceptually, it might be more natural, as it is based on the spectral decomposition of the time evolution operator $e^{-iHt}$, which is a standard way to evolve a wave function. It also brings more insight into the problem---We now understand why $A_0$ is a nonsmooth function of $t$, which is the main point of \cite{epl1}---The coefficients of its Fourier series (\ref{An}) just decay too slowly \cite{zhang}.

\begin{equation*}
  H = - \sum_{m=-\infty}^\infty\left (a_m^\dagger a_{m+1} + a_{m+1}^\dagger a_m + \frac{U}{2} a_m^\dagger a_m^\dagger a_m a_m \right ) + V a_0^\dagger a_0
\end{equation*}

\begin{equation*}
  Z = \sum_{{odd-parity}} e^{-\beta E_n}  + \sum_{{even-parity}} e^{-\beta E_n}   =\int d E (\rho_{odd}(E) + \rho_{even}(E)) e^{-\beta E }
\end{equation*}

\section{Conclusions and discussions}

In conclusion, we have solved analytically the eigenvalues and eigenstates of a toy model, which arose in some quantum quench dynamics problem. This on the one hand fills a gap, and on the other hand provides a different (possibly more natural) approach to studying its dynamics.

The toy model should be of some pedagogical value, as the model itself and the solution presented here are all simple enough. It can serve as a good exercise in a undergraduate  course of quantum mechanics.



\section*{Acknowledgments}

The authors are grateful to J. Guo and K. Yang for their helpful comments. This work is supported by the National Science Foundation of China under Grant No. 11704070.

\end{document}